\documentclass[]{spie}  

\usepackage{amsmath,amsfonts,amssymb}
\usepackage{graphicx}
\usepackage[colorlinks=true, allcolors=blue]{hyperref}
\usepackage{mathtools,etoolbox}
\usepackage{hhline}
\usepackage{multirow}
\usepackage{booktabs}
\usepackage{caption}
\usepackage{subcaption}
\usepackage{svg}
\usepackage{array, caption, floatrow, tabularx, makecell, booktabs}%

\title{Probabilistic 3D Segmentation for Aleatoric Uncertainty Quantification in full 3D Medical Data}

\author[ab]{Christiaan G.A. Viviers}
\author[a]{M.M. Amaan Valiuddin}
\author[a]{Peter H.N. de With}
\author[a]{Fons van der Sommen}

\affil[a]{Eindhoven University of Technology, 5612 AP, Eindhoven, the Netherlands;}
\affil[b]{Philips IGT, 5684 PC, Best, the Netherlands;}

\authorinfo{Further author information: Christiaan G.A. Viviers: E-mail: c.g.a.viviers@tue.nl: Telephone: +31 (0)6 206 60171}

\begin{document} 
\maketitle

\begin{abstract}
Uncertainty quantification in medical images has become an essential addition to segmentation models for practical application in the real world. Although there are valuable developments in accurate uncertainty quantification methods using 2D images and slices of 3D volumes, in clinical practice, the complete 3D volumes (such as CT and MRI scans) are used to evaluate and plan the medical procedure. As a result, the existing 2D methods miss the rich 3D spatial information when resolving the uncertainty. A popular approach for quantifying the ambiguity in the data is to learn a distribution over the possible hypotheses. In recent work, this ambiguity has been modeled to be strictly Gaussian. Normalizing Flows (NFs) are capable of modelling more complex distributions and thus, better fit the embedding space of the data. To this end, we have developed a 3D probabilistic segmentation framework augmented with NFs, to enable capturing the distributions of various complexity. To test the proposed approach, we evaluate the model on the LIDC-IDRI dataset for lung nodule segmentation and quantify the aleatoric uncertainty introduced by the multi-annotator setting and inherent ambiguity in the CT data. Following this approach, we are the first to present a 3D Squared Generalized Energy Distance ($D^2_\text{GED}$) of 0.401 and a high 0.468 Hungarian-matched 3D IoU. The obtained results reveal the value in capturing the 3D uncertainty, using a flexible posterior distribution augmented with a Normalizing Flow. Finally, we present the aleatoric uncertainty in a visual manner with the aim to provide clinicians with additional insight into data ambiguity and facilitating more informed decision-making. Our code is publicly available at: \url{https://github.com/cviviers/prob_3D_segmentation}

\end{abstract}

\keywords{Aleatoric Uncertainty, Segmentation, Normalizing Flows, Volumetric Medical Data }

\section{Introduction}

With the broad acceptance of deep learning-based computer-aided diagnosis and computer-aided detection (CAD) methods, increasingly more requirements are being posed for their successful deployment. These CAD methods typically assist clinicians with decision-making about potentially critical medical procedures or the planning thereof. While most research has focused on maximizing a specific accuracy metric, in practice, a high accuracy along with a strong indication of potential uncertainty or ambiguity in the model output is extremely valuable and highly relevant.
Deep learning-based semantic segmentation methods using convolutional neural networks have successfully been adopted as CAD methods for a wide range of medical imaging modalities. While research has been conducted towards quantifying the types of uncertainty occurring when using a segmentation model, most of this work is limited to the quantification of the uncertainty in two-dimensional slices or images, where the latter often originate from a 3D volume such as in CT and MRI. However, these approaches fail to exploit the rich 3D features that may help in resolving ambiguities in the volume. In this research, we use the full 3D volume as input to derive a reliable estimate of the aleatoric uncertainty in the segmentation output. This approach potentially enables a better 3D visualization of this uncertainty in clinical practice. 

Two types of uncertainty are typically prevalent in deep learning-based image analysis methods: \textit{aleatoric} and \textit{epistemic} uncertainty~\cite{KendallG17}. Aleatoric uncertainty is an estimate of the intrinsic, irreducible ambiguity in data. It is usually associated with inherent noise in the data and its acquisition process. Epistemic uncertainty is the uncertainty about the model, either as a result of the architecture or the true parameter values of the model due to limited knowledge, e.g. a finite training set size. In this work, we employ the LIDC-IDRI lung CT dataset~\cite{lidc}, which makes use of multiple ground-truth annotations per lung nodule. As described in earlier work of Valiuddin~\textit{et al.}~\cite{valiuddin2021improving}, the epistemic uncertainty -- i.e. preferences, experiences and knowledge -- of the annotators manifests into aleatoric uncertainty when providing annotations as ground-truth data. The different annotations per nodule adds ambiguity during training of a segmentation network. During the annotation process, the radiologist typically annotates on a single 2D plane of the 3D volume. However, whilst annotating, full access to the other two views of the CT scan are typically available on the same screen. This allows the annotator to correct the annotation if it does not align with the other two views and as a result, the annotator creates a true 3D annotation. In the LIDC-IDRI dataset, as described in Section~\ref{dataprep}, annotators were allowed multiple rounds of annotation, thereby potentially increasing the quality of the annotation by using the full 3D information available. 

In recent research, various methods have been proposed to quantify the uncertainty arising in segmentation models or resulting from images. One increasingly popular approach is the Probabilistic U-Net~\cite{kohl2018probabilistic}, as proposed by Kohl~\textit{et al.}. They propose combining a 2D U-Net with a conditional variational autoencoder (VAE) capable of learning a distribution over the possible annotations and ultimately construct a generative segmentation model. The Probabilistic U-Net provides compelling results in resolving the ambiguity in an image. More recently, various improvements to this model have been proposed~\cite{selvan2020uncertainty, valiuddin2021improving,hu2019supervised}, paving the way as one of the leading uncertainty quantification methods. Selvan~\textit{et al.}~\cite{selvan2020uncertainty} and Valiuddin~\textit{et al.}~\cite{valiuddin2021improving} propose adding a Normalizing Flow to the posterior network of the Probabilistic U-Net. This allows the model to move away from modelling the ambiguity as strictly axis-aligned Gaussian and, instead, allows for a learned posterior distribution of varying complexity. 

This work provides the following contributions. First, we propose a 3D probabilistic framework, which builds upon the research from both Kohl~\textit{et al.} and Valiuddin~\textit{et al.} that exploits the full-3D spatial information to resolve the uncertainty in the original CT volumes. Second, it is shown that more diverse segmentations are obtained when the posterior distribution is enhanced by a Normalizing Flow. This finding suggests that such modeling enables capturing the uncertainty more accurately. Third, we test the proposed method's ability to capture uncertainty on the LIDC-IDRI lung nodule datatset and are the first to present results in the 3D version of the $D^2_\text{GED}$ metric, and show that a high segmentation accuracy is obtained using a Hungarian-matched 3D IoU. 

In this work, we present a 3D probabilistic segmentation model that exploits the full 3D-spatial information to more accurately quantify the aleatoric uncertainty, while maintaining 3D consistency. The proposed model is equipped with a Normalizing Flow to eliminate the strictly Gaussian latent space that is currently enforced to further improve the model's ability to quantify the uncertainty. More specifically, the model consists of a 3D U-Net and a 3D conditional VAE enhanced with the Normalizing Flow, to generate a diverse set of plausible segmentations.
\section{Related Work}
The Probabilistic U-Net~\cite{kohl2018probabilistic} is capable of generating a diverse set of valid segmentation hypotheses. This model was proposed by Kohl~\textit{et al.} as a method for capturing the ambiguity in an image. Shi Hu~\textit{et al.}~\cite{hu2019supervised} showed how this ambiguity can be interpreted as uncertainty. They propose adding variational dropout and an additional inter-grader variability term to the training objective of the Probabilistic U-Net. This change allows the model to capture a combination of the epistemic and aleatoric uncertainty. Selvan~\textit{et al.}~\cite{selvan2020uncertainty} improve the quantification of aleatoric uncertainty by adding a Normalizing Flow to the posterior network of the VAE~\cite{rezende2015variational} in the Probabilistic U-Net.  Valiuddin~\textit{et al.}~\cite{valiuddin2021improving} indicate that an additional metric, the Hungarian-matched IoU, is necessary to effectively evaluate the performance of these uncertainty quantification methods. 

The aforementioned studies have all been conducted with a focus on 2D images and slices from the original 3D volume, which is a step away from the domain where the true uncertainty resides. In addition, these methods all heavily focus on a specially crafted subset of the LIDC-IDRI dataset. The image patches used to train and test the models exactly contain the regions where the ambiguity resides, and nothing more. In practice, at test time, this is not the case. In this work we use a fixed 3D volume surrounding the lung nodules and consider all the image patches, with or without uncertainty.  



\section{Methods}

\subsection{Model Architecture}
\begin{figure}
\begin{minipage}{\linewidth}
  \centering  \centerline{\includegraphics[width=0.9\linewidth]{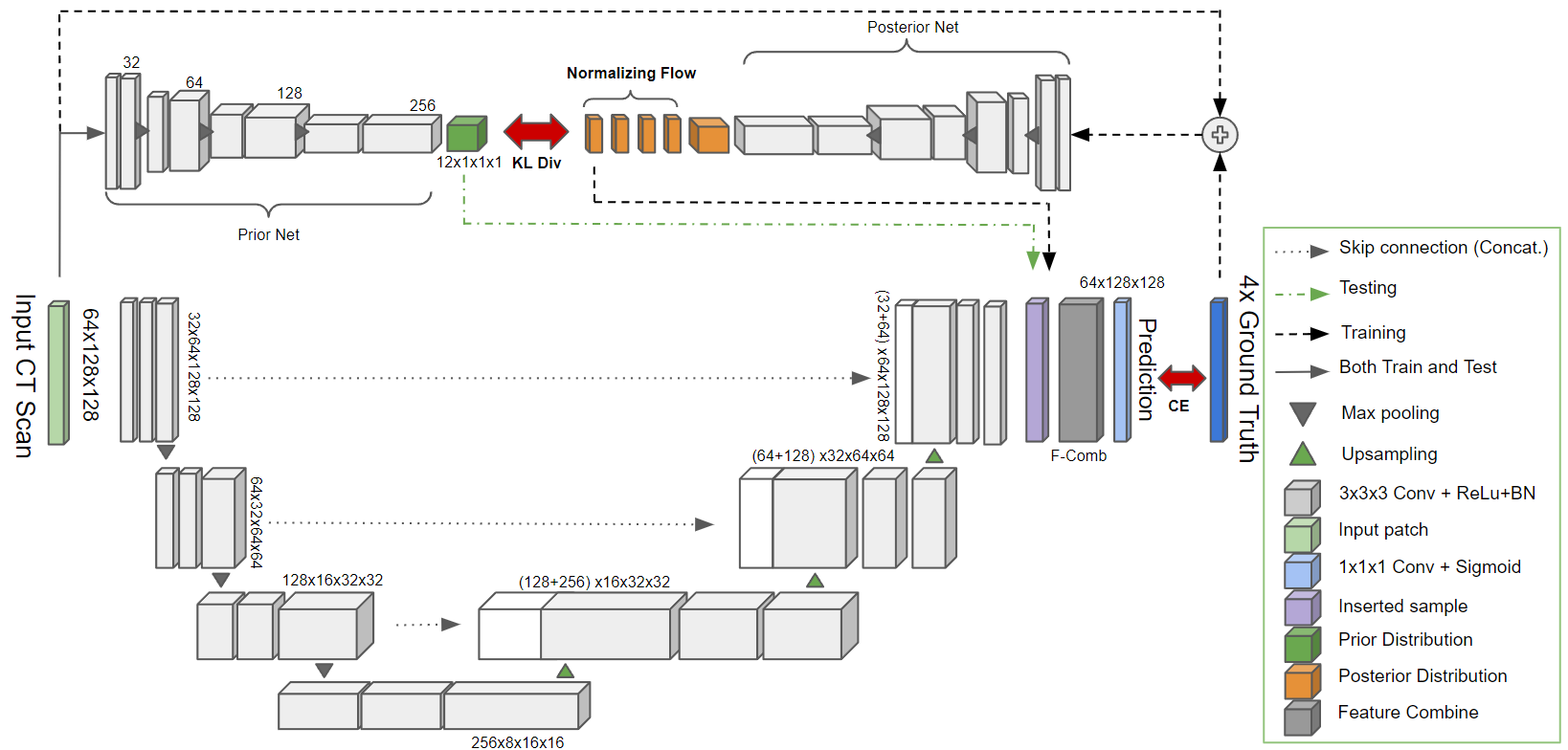}}
  \caption{\textit{Diagram of the 3D Probabilistic U-Net with an augmented flow posterior. The bottom network depicts the 3D U-Net, the Prior and Posterior network are shown at the top and a Feature combination network at the right combines samples taken from the captured distributions.  The diagram additionally depicts both the training and testing configuration.}}
  \label{fig:arch}
\end{minipage}
\end{figure}

In this research, we extend the Probabilistic U-Net to the 3D domain and address a key limitation by augmenting the posterior network with a Normalizing Flow (NF). The network consists of a 3D U-Net, a Prior network, Posterior network enhanced with an NF, and a Feature Combination network. By combining the 3D spatial features extracted by the U-Net with samples taken from a latent distribution encapsulating the solution space, a set of diverse, but plausible segmentations can be generated. The standard deviation across these predictions can be interpreted as the aleatoric uncertainty. The U-Net~\cite{unet} and 3D U-Net~\cite{3dunet} have shown time and again their ability to segment structures of interest at state-of-the-art performance. While we employ the 3D U-Net to obtain the relevant 3D spatial information, this approach is generic and allows for any other segmentation network to be used. A deep CNN conditioned on the input CT scan is used to model a low-dimensional axis-aligned Gaussian latent space, representing the segmentation variants (Prior distribution). Another CNN-based axis-aligned Gaussian encoder (Posterior network) that is conditioned on both the query CT scan and a ground-truth segmentation, is utilized during training to model a posterior low-dimensional latent space. Valiuddin~\textit{et al.} point out the shortcomings in modelling the posterior distribution to be strictly Gaussian. As such, we augment our posterior network with either a 2-step planar or radial flow~\cite{Kobyzev_2020}, to potentially increase the complexity of the captured posterior distribution, and thereby provide more meaningful updates to the prior network during training. Figure~\ref{fig:arch} portrays a detailed diagram of the proposed network architecture.
During training, the Probabilistic 3D U-Net makes use of the Posterior network, Prior network, U-Net and the feature combination layers. Samples are taken from the image-label conditional distribution captured by the Posterior network and combined with the features extracted from the U-Net through the Feature Combination network. The loss is then computed using Equation~(\ref{lossfunc}). The Prior network follows the Posterior network during training, as enforced by the KL-divergence, and thus learns to capture this image-label conditional distribution from the image alone. At test time, the Posterior network is discarded and samples are taken from the Prior network instead. It should be noted that there is only one forward pass through the U-Net (for image feature extraction) and the Prior network (to capture the image-conditional distribution). However, multiple passes through the feature combination network are made, in order to combine a new sample from the Prior distribution with the image features.     
\subsection{Loss Function \& Evaluation Criteria}
In line with previous work on conditional variational autoencoders, our training objective consists of minimizing the variational lower bound \cite{kingma2014autoencoding}. This entails minimizing a cross-entropy difference (in our case) between the ground-truth segmentation ($\textbf{y}$) and a prediction ($\textbf{s}$), minimizing the Kullback-Leibler~($\operatorname{KL}$) divergence between the posterior distribution ($p_{\phi}$) and the prior distribution ($p_{\theta}$), and finally, a correction term for the density transformation through the Normalizing Flow~\cite{berg2019sylvester}. Given a query image ($\mathbf{x}$) and a posterior sample ($\mathbf{z}$), $p_{\psi}$ (a U-Net and Feature combination network) generates a plausible segmentation ($\textbf{s}$). This loss term can formally be specified by
\footnotesize
\begin{equation}
    \begin{aligned}
        \mathcal{L}(\mathbf{y}, \mathbf{x}, \theta, \phi, \psi) &= - \, \mathbb{E}_{p_\phi(\mathbf{z}\vert\mathbf{y},\mathbf{x})}[\operatorname{log}p_{\psi}(\mathbf{y}\vert\mathbf{z},\mathbf{x})]
        + \beta \cdot \left(\operatorname{KL}\left(\,p_{\phi}(\mathbf{z}_0\vert\mathbf{y},\mathbf{x})\vert\vert p_{\theta}(\mathbf{z}\vert\mathbf{x})\,\right) -\mathbb{E}_{p_\phi(\mathbf{z}_0\vert\mathbf{y},\mathbf{x})}\left[\sum_{i=1}^{K} \log\left( \left|\operatorname{det} \frac{d f_{i}}{d \mathbf{z}_{i-1}}\right|\right)\right]\right).
    \end{aligned}
    \label{lossfunc}
\end{equation}
\normalsize
These losses are combined and weighed using hyperparameter $\beta$~\cite{fu2019cyclical, alemi2018fixing}. The original research provides a detailed derivation of the elbo loss~\cite{kingma2014autoencoding}, how it is used in the context of Probabilistic U-Net~\cite{kohl2018probabilistic} and the NF-likelihood objective~\cite{Kobyzev_2020,valiuddin2021improving, berg2019sylvester}. 

\noindent The metric $D^2_\text{GED}$ or Squared Generalized Energy Distance has become the \textit{de-facto} metric in the context of uncertainty quantification and the quantification of the distance between distributions of segmentations. This metric is defined as
\begin{equation}\label{ged}
    \begin{aligned}
        D^2_\text{GED}(P_\text{GT}, P_\text{Out})=2\mathbb{E} \left[ d(\mathbb{S},\mathbb{Y}) \right] -\mathbb{E}\left[d(\mathbb{S},\mathbb{S}')\right]-\mathbb{E}\left[d(\mathbb{Y},\mathbb{Y}')\right],
    \end{aligned}
\end{equation}
where $d$ is a distance measure, $1 - \text{IoU}(x, y)$, in our case. The parameters $\mathbb{S}$ and $\mathbb{S}'$ are independent samples from the predicted distribution $P_\text{Out}$. The parameters $\mathbb{Y}$ and $\mathbb{Y}'$ are the 4 samples from {the ground-truth distribution $P_\text{GT}$. In addition to the $D^2_\text{GED}$, we also report the Hungarian-matched IoU. This compensates for a shortcoming in the $D^2_\text{GED}$ that when the predictions are relatively poor, the metric rewards sample diversity by definition. We duplicate the ground-truth set (4 annotations) to match the desired sample number when computing the Hungarian-matched IoU. This measure calculates the distance between two discrete distributions by determining an optimal coupling between the ground-truth and prediction set subject to the IoU metric. 

\subsection{Dataset \& Data Preparation}\label{dataprep}

To evaluate the proposed method's ability to resolve the ambiguity in the data, we use the popular LIDC-IDRI dataset. This dataset contains the lung CT scans from 1,010~patients with manual lesion annotations from up to 4~experts. In total, there are 1,018~CT scans potentially containing multiple lung nodules of different levels of malignancy. In this work, we have used the annotations from a second reading, in which the radiologists were presented an anonymized version of the annotations from other experts and were allowed to make adjustments to their own annotations. Contrary to previous work, we use every nodule in the dataset if it has been annotated by at least one radiologist (potentially missed by three), regardless of the shape or severity of the nodule. We pre-process the CT scans by clustering all nodule annotations for a scan, by computing a distance measure between the annotations. If an annotation is within one voxel spacing of that particular CT scan from another annotation, it is grouped to belong to the same nodule. The scan is resampled to a 0.5~mm along the $x$ and $y$-dimensions and 1~mm along the $z$-dimension. This is followed by cropping the CT scan and resulting annotations based on the center of mass of the first annotator's mask with a dimension of 96$\times$180$\times$180 voxels in the $z, x, y$-dimensions. Finally, if the nodule does not have at least 4~annotations, the ground-truth (GT) masks are filled with empty annotations. This addition is made to be consistent with previous work on this dataset~\cite{kohl2018probabilistic, valiuddin2021improving} and to capture the difficulty in detecting a nodule. This results in a total of 2,651~3D patches, each containing a nodule and four annotations. An example of the nodule in the CT scan and the four ground-truth annotations are depicted in Figure~\ref{fig:example masks}.


\subsection{Experiments}
To compare the proposed approach against prior work, we conduct six experiments. We train the (1)~original Probabilistic 2D U-Net and the (2)~Radial NF-augmented Probabilistic 2D U-Net on 2D axial slices of the 3D volume. In practice, we filter the slices based on the presence of at least one positive annotation from any of the raters and use them for training, to avoid a heavily imbalanced training set. The (3)~3D U-Net, (4)~Probabilistic 3D U-Net and an (5-6)~NF-augmented (Radial and Planar) Probabilistic 3D U-Net are then trained on the 3D patches. In contrast to prior work where the 3D lesion was sliced and split into 2D images, where some 2D slices potentially land in the training set and some in the validation/test set, we conduct our experiments on a per-lesion basis. This avoids any potential model bias caused by the splitting and makes the proposed approach more clinically relevant, since we can present the uncertainty for each lesion.
\begin{figure}[H]
\begin{minipage}{\linewidth}
  \centering  \centerline{\includegraphics[width=0.80\linewidth]{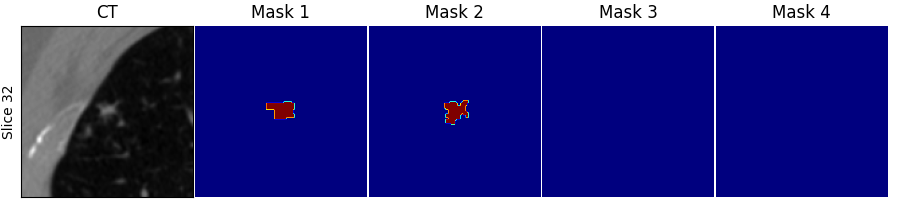}}
  \caption{\textit{Example nodule in a slice from the CT scan and the four ground-truth annotations.}}
  \label{fig:example masks}
\end{minipage}
\end{figure}
In our implementation, we split the nodule data in a 70/15/15 training \& validation/test split. During training, we randomly sample one of the four annotations to be used as ground-truth segmentation and crop the CT volume and label to 64$\times$128$\times$128 voxels. In line with previous work, for the 3D Probabilistic U-Net, the dimensionality of the latent space is set to $L=6$. The proposed framework is implemented in PyTorch and extends on the work conducted by Wolny~\textit{et al.}~\cite{eLife}. We train using a batch size of~32 in the 2D case and 4~in the 3D case. An Adam optimizer with an initial learning rate of $1\times 10^{-4}$ and a weight decay of $1\times 10^{-5}$ is used. We reduce the learning rate by a factor of~0.2 if the validation loss does not decrease after 20~epochs. The $\beta$ parameter is controlled using a cosine cyclical annealing strategy as descibed by Fu~\textit{et al.}~\cite{fu2019cyclical}. In all our 3D Probabilistic U-Net experiments, we use the same hyperparameters and a hardware configuration with an RTX~3090Ti GPU (available from Nvidia Inc. Santa Clara, CA, USA). Training to completion takes about 2 days on the average. For performance evaluations, we report results using a more readily available RTX~2080Ti GPU.
\section{Results}\label{sec:results}
The results of our experiments are shown in Figure~\ref{fig:resuls_all_models}, Figure~\ref{fig:2Dvs3D}, Figure~\ref{fig:one_nodule} and Table~\ref{tab:results}. In Figure~\ref{fig:resuls_all_models}, example predictions from all the models used in our experiments are showcased for qualitative evaluation. Here, \textbf{$\mu$}~GT refers to the mean segmentation of the four raters and \textbf{$\mu$}~Pred is the mean of the predictions. This mean prediction is the segmentation recommended by the Probabilistic 3D U-Net. Additionally, the figure depicts the variation in the segmentations. More specifically, the standard deviation of the ground-truth labels (\textbf{$\sigma$}~GT) and the logits (after sigmoid activation) resulting from the model predictions (\textbf{$\sigma$}~Pred) are depicted. In the figure it can be observed that this deviation across the predictions can be interpreted as the uncertainty. We scale the uncertainty heat map visualization to the the maximum standard deviation of a particular model's predictions. Additionally, the figure depicts a rather conservative segmentation of part of the lesion from the deterministic 3D U-Net segments, while the other models are capable of producing a more accurate segmentation. 
\begin{figure}[H]
\begin{minipage}{\linewidth}
  \centering  \centerline{\includegraphics[width=0.80\linewidth]{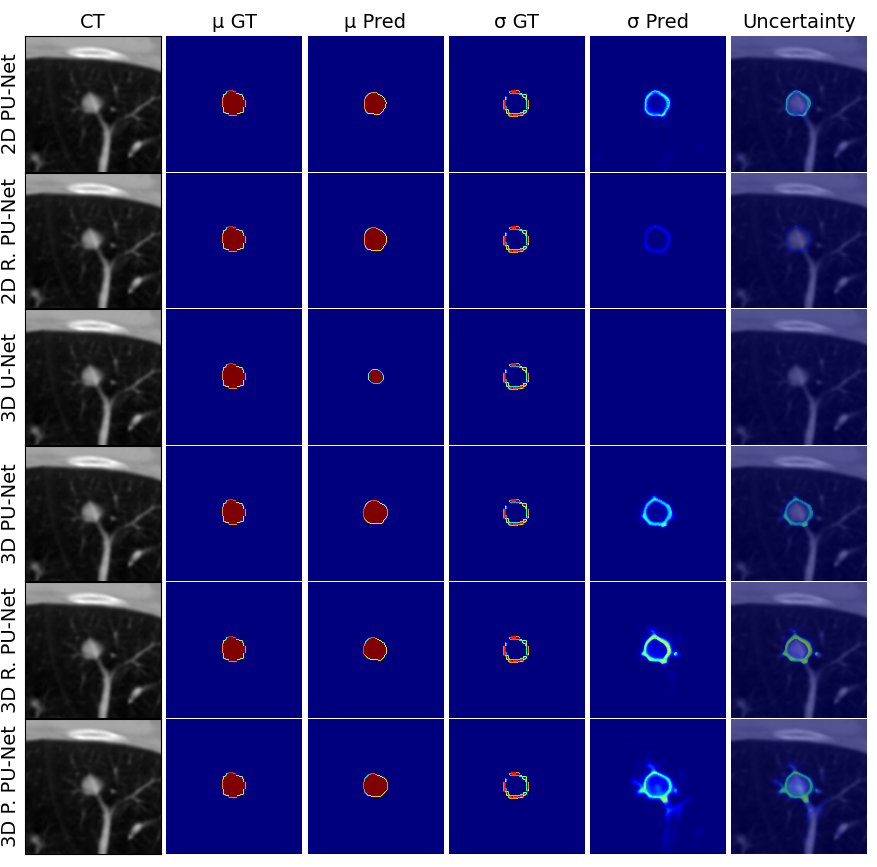}}
  \caption{\textit{Example predictions for the same data slice with a nodule from the 3D U-Net, 2D \& 3D Prob.U-Net in the test set. This same slice is used to get a comparative sense of model performance.}}
  \label{fig:resuls_all_models}
\end{minipage}
\end{figure}
Figure~\ref{fig:2Dvs3D} depicts the predictions for the 2D and 3D Prob.U-Net for 2~slices from a nodule in the test set. It can be seen that the 2D model misses the nodule in Slice~34, while its 3D counterpart correctly detects it. Although the 2D model has some uncertainty about the presence of the nodule, it is rather low.  
\begin{figure}[H]
\begin{minipage}{\linewidth}
  \centering  \centerline{\includegraphics[width=0.80\linewidth]{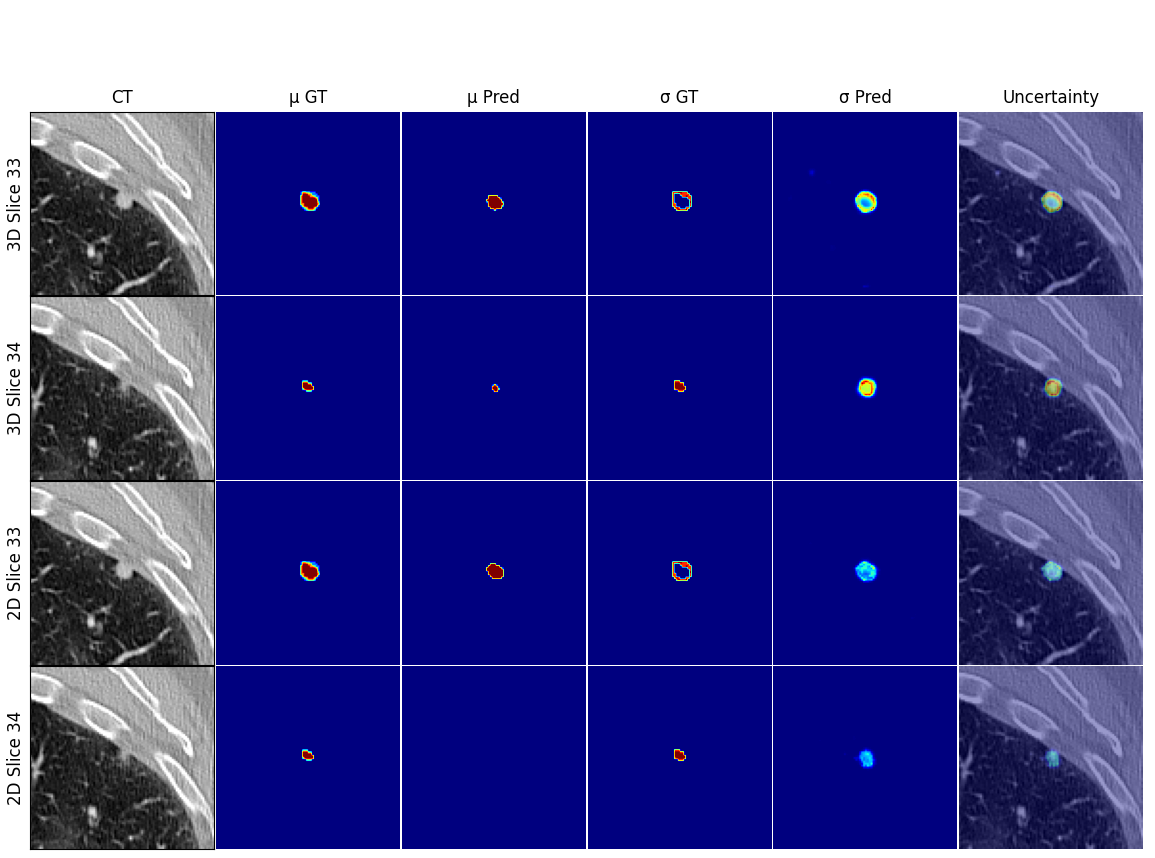}}
  \caption{\textit{Zoomed example 3D \& 2D Prob.U-Net predictions for 2 slices from a nodule in the test set.}}
  \label{fig:2Dvs3D}
\end{minipage}
\end{figure}
In Figure~\ref{fig:one_nodule} multiple consecutive slices are depicted of a CT scan from our test set and Prob. 3D U-Net predictions for a nodule. Slice~25 displays some uncertainty from the model about the presence of a lesion although no rater indicated its existence yet. In the next slice the lesion is clearly delineated by the raters and the model captures the uncertainty in a similar fashion as the disagreement between them. Slices~27-31 and~33-35 are not shown, since the model correctly segments and captures the uncertainty in comparison to the raters. Slice~38 reveals the large lesion as shown by the annotations from the raters, but it rapidly disappears towards Slice~39. However, the model still segments the lesion in Slice~39 and expresses high uncertainty. 
\begin{figure}[H]
\begin{minipage}{\linewidth}
  \centering  \centerline{\includegraphics[width=0.80\linewidth]{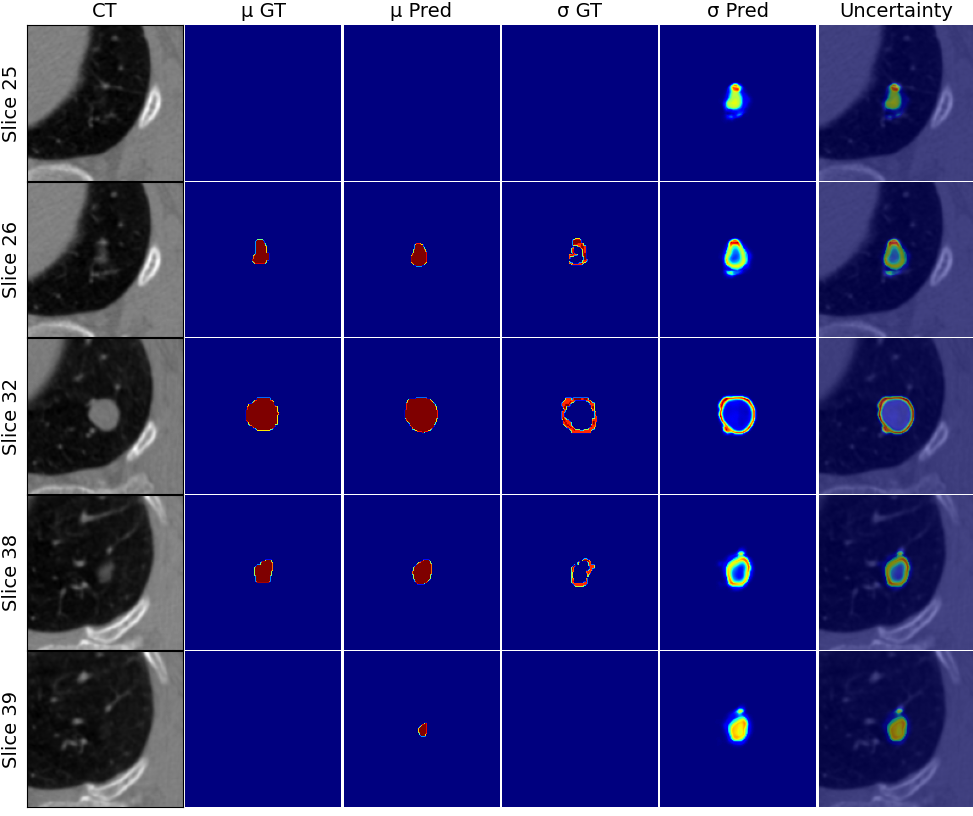}}
  \caption{\textit{Example 3D Prob.U-Net predictions for a multiple slices from a nodule in the test set.}}
  \label{fig:one_nodule}
\end{minipage}
\end{figure}
We quantitatively compare the proposed approach with the 2D counterparts, aiming to resolve the ambiguity in the LIDC-IDRI dataset in Table~\ref{tab:results}. We compute the 2D $D^2_\text{GED}$ and 2D Hungarian IoU on a per-slice basis and take the average across all the slices of the lesion, ignoring slices with empty ground-truths and predictions. The 3D $D^2_\text{GED}$ and 3D Hungarian IoU are immediately computed on a per-case level and then averaged across the test set. In the case of the 2D models, during the forward pass of a single 2D slice, an image conditional prior distribution is computed. We then draw 16~samples from this distribution. For the next slice in the series of the lesion, a completely new prior distribution is presented at inference time. As such, the uncertainty captured by this prior distribution is inconsistent over the individual slices of the lesion and it is not possible to reconstruct a consistent and true 3D segmentation using this approach (certainly not with 3D metrics such as 3D $D^2_\text{GED}$ and 3D IoU calculation). For the 3D Probabilistic U-Net, we report the 2D Hungarian-matched IoU and 2D $D^2_\text{GED}$ (2D IoU distance averaged along the $z$-axis) and the 3D Hungarian-matched IoU and $D^2_\text{GED}$.
\begin{table}[H]
\centering
\begin{tabular}{l| cc|cc} 
\toprule
     Model  & {2D$\downarrow$ $D^2_\text{GED}$} & {2D$\uparrow$ IoU}  & {3D$\downarrow$ $D^2_\text{GED}$}  &  {3D$\uparrow$ IoU} \\ 
    \midrule
     Kohl~\textit{et al.}   & 0.445 \enspace &  0.473 \enspace & N/A \enspace  & N/A \\
     Valiuddin~\textit{et al.}  & 0.441 \enspace &  0.481 \enspace & N/A \enspace  & N/A \\
     3D U-Net  & 1.283 \enspace &  0.332 \enspace & 1.263 \enspace  & 0.383 \\
    \midrule
     3D Prob.U-Net  & 0.427 \enspace & 0.510 \enspace  & 0.422  \enspace  & 0.457\\
    \hspace{0.1cm} + Planar Flow\enspace  & \textbf{0.417} \enspace  & 0.511 \enspace & \textbf{0.393}   \enspace  & 0.465 \\
    \hspace{0.1cm}  + Radial Flow\enspace  & 0.429 \enspace  & \textbf{0.520} \enspace & 0.401 \enspace  &  \textbf{0.468}  \\
    \bottomrule
\end{tabular}
\caption{\textit{Evaluations on the LIDC-IDRI test set (15\%) of the different methods on the $D^2_\text{GED}$ and Hungarian IoU metric based on 16 samples.} \label{tab:results} }
\end{table}

We compare the inference time of the 3D U-Net and the Prob. 2D and 3D U-Net per nodule volume ($64\times128\times128$ voxels). We do not include additional results on the models with NF-augmented posteriors networks, since the 2-step low-dimensional bijective transformation has a negligible computational time footprint in comparison to the rest of the network. Table~\ref{tab:operation} showcases the computation time per operation for the different models and with different batch sizes (BS). It can be seen that the Prob. 3D U-net has a shorter inference time for a volume of this size, compared to its 2D counterpart. 
\begin{table}[H]
\centering
\begin{tabular}{l|cccc} 
\toprule
     Operation &  3D U-Net(BS 1) & 2D PU-Net(BS 1) & 2D PU-Net(BS 64) & 3D PU-Net(BS 1)   \\ 
    \midrule
    Forward pass    & 2.34 (ms) \enspace & 5.75 (ms) $\times 64$\enspace &  397.27 (ms) \enspace & 124.31 (ms) \enspace   \\
     Sample + F-comb ($\times 1$) &  N/A \enspace  & 0.51 (ms) $\times 64 \times 16$ \enspace & 157.09 (ms)$\times 16$  \enspace & 8.44 (ms)$\times 16$ \enspace   \\
    \midrule
     Total  & 2.34 (ms) \enspace & 892.29 (ms)\enspace & 2910.51 (ms)\enspace & 259.35 (ms) \\
    \bottomrule
\end{tabular}
\caption{\textit{Inference time (per operation) of the 3D U-Net and 16 samples from the Probabilistic 2D and 3D U-Net per nodule ($64\times128\times128$ voxels). BS is the Batch Size.} \label{tab:operation} }
\end{table}

\section{Discussion}
This research extends the Probabilistic U-Net to the 3D domain to utilize the rich 3D spatial information when resolving the uncertainty. We introduce the Probabilistic 3D U-Net and employ recent improvements in the 2D Probabilistic U-Net, by adding either a planar or radial flow to the posterior network. This augmentation with NFs enables capturing distributions of various complexity, thereby relaxing the strictly axis-aligned Gaussian constraint previously employed. To test the model's ability to capture the aleatoric uncertainty, we use the LIDC-IDRI dataset for benchmark tests.

Section~\ref{sec:results} displays the results of the conducted experiments. For qualitative evaluations, Figure \ref{fig:resuls_all_models} showcases example predictions from all the models used in this research. It can be noted that all of the models perform well on this clearly defined lesions, except for the 3D U-Net. The 3D U-Net segments the lesion in a conservative manner, possibly due to seeing many empty ground-truth labels and not being able to capture this ambiguity in a meaningful way. The Prob. 3D U-Net with planar flow expresses more uncertainty about various parts of the lesion. Figure~\ref{fig:2Dvs3D} highlights example predictions of the Prob. 3D U-Net where information from the complete 3D volume is used to detect, segment and resolve the uncertainty about the lesion in axial Slice~34 of the CT scan. The same nodule is missed by the Prob. 2D U-Net, due to a lack of information from prior slices. These results are also reflected in the 2D $D^2_\text{GED}$ and Hungarian IoU, as shown in Table~\ref{tab:results}, with the 3D models outperforming the 2D models. Interestingly, in Figure~\ref{fig:one_nodule}, the 3D spatial awareness of the model is showcased through the uncertainty expressed in Slice~25. A rather large lesion is coming up (iterating through the CT slices in an ascending order) and the model expresses uncertainty about the exact starting position, since the raters segmented no lesion followed by a large lesion in consecutive slices. The same phenomenon can be seen moving from Slice~38 to~39, although here the model incorrectly (according to the raters) segments the lesion while it is still partially visible.

By the performance evaluations presented in Table~\ref{tab:operation}, it can be observed that it is the most efficient to present the uncertainty with a 3D Prob. U-Net. For a volume of 64$\times$128$\times$128 voxels, the 64~2D slices can be passed through the Prob. 2D U-Net in a large batch, but it scales poorly in comparison to a single-slice forward pass. The forward pass of the Prob. 2D U-Net with a batch size of 64~takes 397.27~ms to compute. Drawing 16 samples from the Prior distribution and combining it with the 2D U-Net features through the Feature combination networks takes 2513.44~ms (157.09~ms$\times$16). In total this approach will take 2910.51~ms to compute compared to the approximate 10$\times$ speed improvement that is required for calculating the uncertainty with the Prob. 3D U-Net (259.35~ms). It should be noted that significant computational time drawbacks occur when using the Prob. 3D U-Net in comparison to the the standard 3D U-Net (2.34~ms for inference), although this is not a realistic alternative since no uncertainty can be expressed.  

\section{Conclusion}
In CAD methods, it is important to provide clinicians with an accurate measure of uncertainty when they evaluate and plan their procedures. Accurately capturing and presenting segmentation uncertainty will increase clinician confidence in model predictions and facilitates better informed decision-making. Existing CT-based segmentation methods aim to do so by quantifying the uncertainty from 2D image slices, whereas the true uncertainty resides in the full 3D CT or MRI volume. We propose a novel 3D probabilistic segmentation model that is capable of resolving and presenting the aleatoric uncertainty in 3D volumes through diverse and plausible nodule segmentations. The model consists of a Deep 3D U-Net and a 3D conditional VAE that is augmented with an Normalizing Flow (NF) in the posterior network. NFs allow for more flexible distribution modelling and as such, we alleviate the strictly Gaussian posterior distribution that was previously enforced. We test our approach on the LIDC-IDRI lung nodule CT dataset. This is, to the best of our knowledge, the first approach that presents the 3D Squared Generalized Energy Distance ($D^2_\text{GED}$) and 3D Hungarian-matched IoU for lung nodule segmentation and uncertainty prediction. We quantify the uncertainty prediction performance and achieve a 0.401 3D $D^2_\text{GED}$ and a Hungarian-matched 3D IoU of 0.468 with the radial Prob. 3D U-Net. In addition, since the model uses the full native 3D volumes, it is a step closer to the practical application of accurately segmenting and presenting uncertainty in 3D CT data. Finally, we present the aleatoric uncertainty, computed as the standard deviation across the model predictions, in a visual manner. This enables an interpretable expression of the uncertainty and is potentially providing clinicians additional insight into data ambiguity and allowing for more informed decision-making.

\bibliography{main} 
\bibliographystyle{spiebib} 

\end{document}